PREPRINT

# Students' Perception Accuracy of Partners' AI Use and its Relation to Collaboration Performance

Laura Graf[1], Ramona Beinstingel[1], Stephan Krusche[1], Oleksandra Poquet[1]

[1] Technical University of Munich, Munich, Germany
{l.graf, ramona.beinstingel, krusche, sasha.poquet}@tum.de

**Abstract.** Collaborative assignments are a cornerstone of programming education. Effective collaboration during a programming project depends on the formation of reasonably accurate beliefs about how each partner works. Generative AI tools, now widely used by undergraduate students, have introduced a consequential and largely invisible new dimension into collaboration: each student's use of AI. When partners collaborate remotely, they interpret partners' ability and effort through their code. This raises the question of how accurately students perceive each other's AI use in collaborations, and if a misalignment in these perceptions relates to team performance. To address this question, we conducted a three-wave longitudinal study of 103 student pairs in an introductory software engineering course. We found that greater misalignment between partners' beliefs about each other's AI use early in the project was associated with lower final project scores. The effect of such misaligned perceptions is the strongest in teams with lower prior programming performance, suggesting that low performing students pay a higher cost of misaligned perceptions. The perception misalignment does not consistently decrease through face-to-face pair-programming sessions. This suggests that ways to foster transparency may be needed to support student teams in collaborative programming.

**Keywords:** Generative AI · Collaboration· Programming education · Perception accuracy · Team performance

## 1 Introduction

Collaborative programming projects are a cornerstone of programming education. They develop students' technical proficiency and communication skills required in professional software development [1]. In team-based programming assignments, students work on a shared codebase and must divide tasks, review each other's code, and, among other tasks, align on design decisions. These are all examples of coordination activities that are central to effective collaboration [2]. To coordinate effectively, the students need to be aware of how their collaborators work, what they know, and how reliably they produce code that others can extend [3]. From that standpoint, it is plausible that failure to accurately judge partners' AI use in a collaborative task may disrupt coordination and potentially influence collaboration outcomes.

Generative AI tools challenge coordination, since AI use is largely invisible to the other team members. Traditionally, team members form expectations and beliefs about their peers by observing visible traces that reflect partner's work. For instance, in programming teams partners infer each other's practices and abilities from their commit patterns, code structure, testing habits, and the pace and volume of code production [4]. These traces provide cues for interpreting each other's contributions and adjusting coordination. With many students now relying on large language models and AI-assisted coding tools during programming tasks, there is a shift in what their partners can infer about them during the process. A large volume of well-structured code may reflect either strong individual expertise or extensive AI assistance. However, distinguishing between these possibilities may be difficult [5]. Therefore, collaboration may be indirectly shaped by how aware team partners are of each other's AI use.

Despite its relevance, the consequences of unobserved AI use on collaboration have received little attention. There is a gap in empirical work on team partners' awareness of each other's AI use for learning and work and its implications for collaboration. To address this gap, we conducted a longitudinal study of student teams over the course of a one-month programming project. Three times during one month, we asked 103 student pairs to self-report own and estimate partner AI usage amount and AI usage type. Using these data, we examine (1) how aware students were of each other's AI use over the course of a collaborative programming project, and (2) whether misalignments in these perceptions were negatively associated with team performance. Our results corroborate that some students stayed unaware of their partners' AI use and this affected collaboration outcomes. The findings suggest a need for tools that make partners' AI use visible to the team to improve collaboration.

## 2 Background

Research in group awareness demonstrates that collaboration depends on partners using visible cues in each other's work to form expectations about relevant aspects of their collaborators [6]. These visible cues enable group awareness that supports coordination of their joint activity [6]. The increasing use of AI tools affects how these observable cues can be interpreted, since collaboration aspects inferred from the cues, such as partner's competence, can no longer be trusted [5]. As a result, collaborators' awareness of each other may become limited, negatively affecting collaboration. This raises several related questions: whether collaborators are aware of each other's AI use, how the awareness of AI use develops during collaboration, and whether differences in this awareness are associated with collaboration outcomes.

### 2.1 Team Partners' Awareness of AI Use

To understand how collaborators form awareness of each other, prior research in computer-supported collaborative learning (CSCL) has examined how group awareness is maintained and supported. Within this literature, group awareness refers to the knowledge of aspects relevant for coordination, such as group members' activities, expertise, or level of participation. A central assumption in this work is that these aspects are not directly observed but inferred by group members from observable indicators of collaborators' work [7]. Learning analytics (LA) systems can make indicators of collaboration visible by aggregating and presenting contributions to help team members interpret each other's activity and thereby support coordination [8], [9]. The effectiveness of such an inference, however, rests on a direct link of the observable indicators to the underlying activity or expertise.

The increasing use of generative AI in programming affects how contributions can be interpreted. Because the same observable contributions can plausibly reflect either a collaborator's own work or AI generation [5], a partner can no longer read the sophistication or volume of code as a signal of the other's expertise or effort. From an awareness perspective, this weakens the link between observable partner behaviour and the inferences made from it. For instance, if partners underestimate or overestimate each other's reliance on AI, they may also develop inaccurate expectations about their partner's independent capabilities. The expectations about partners' capabilities affect coordination decisions, such as task allocation, review intensity, or knowing how much support a teammate may require [10].

Prior empirical work on generative AI in student programming has focused primarily on individual use patterns [11], learning outcomes [12], and academic integrity concerns [11]. Research that does address collaboration has primarily examined how AI can be used to mediate and support group processes [13], [14]. What remains underexplored is the situation where multiple collaborators independently use AI as a private support tool. This is a common context in both education and professional contexts, and one that poses challenges for collaborative work. Individuals who misjudge their partner's AI reliance may misallocate tasks, incorrectly conclude how much to scrutinise peer contributions, or fail to offer support where needed. Understanding how awareness of a partner's AI use develops over collaboration and whether misalignment in such awareness affects team performance is therefore consequential for both programming education and the design of tools that support collaborative work.

### 2.2 Formation of Beliefs about Team Partners' AI Use.

Since observable contributions no longer provide reliable cues about how collaborators work, it becomes important to consider how students form beliefs about their partners under such uncertainty. In the absence of direct information about AI use, collaborators must find other ways to estimate how their partner approaches programming tasks. Research on interpersonal perception suggests that when information about others is limited or ambiguous, individuals rely on heuristic strategies to form judgments [15], [16].

One heuristic strategy based on which individuals form beliefs about others is social projection, i.e. individuals assume partners to be similar to them and use their own behaviour as a reference point for judging others [15]. This has been shown in research where students who engage in a particular behaviour (e.g., cheating) are more likely to believe that behaviour is common among peers [17]. From this perspective, students' own AI use may serve as an implicit baseline they use to estimate their partner's use.

Another common heuristic explaining how individuals form beliefs about others under observability constraints is their normative expectation, that is, their assumptions about what behaviour is 'normal', i.e., typical or appropriate in a given context [16]. Applied to AI use, this mechanism suggests that students may have ideas of the 'typical' use of generative AI tools in programming and transfer them to a partner independent of that partner's actual behaviour.

To summarise, the literature suggests that students may estimate their partner's AI use based on their own usage or expectations about the typical use of AI across programming tasks, especially in cases with limited direct communication or observable information about the specific individual. When estimations of partner AI use, and consequently their inferred characteristics, are based on assumed similarity or normative expectations, student judgements of each other would be misaligned among the students with dissimilar AI use. To develop pedagogical interventions supporting student teams, it is therefore necessary to examine estimations of partner AI use and if collaborators develop partner-specific knowledge over time.

### 2.3 Impact of AI Use Perception Misalignment on Collaboration Outcomes

Group awareness is closely linked to how effectively teams coordinate their work. Several distinct theoretical traditions support this claim. Although these traditions emphasise different mechanisms, they agree about the importance of aligned understanding of team partners, their process, and their individual expertise. Shared mental models theory suggests that teams perform more effectively when members hold aligned understandings of relevant aspects of the task and of each other, enabling them to anticipate each other's actions and coordinate with fewer explicit exchanges [3]. Similarly, transactive memory theory emphasises the importance of knowing how knowledge and expertise are distributed within a team, allowing members to rely on each other appropriately and allocate work efficiently [18], [10]. Overall, these theories about collaboration underline that teams coordinate more effectively when members are aware of each other's knowledge, capabilities, and ways of working.

Research is yet to empirically examine if misaligned perceptions of AI use are associated with team performance. Although perception misalignment is likely to influence coordination, its practical relevance depends on its further association with collaboration outcomes. Hence, establishing this association is a necessary first step of investigation.

## 3 Research Questions

We have argued that students' independent use of AI during collaboration activities may challenge group awareness essential for coordination between them. Hence, it is necessary to examine student awareness of their team partners' AI use, and whether misalignment in collaborators' perceptions of each other's AI use is associated with team performance. Accordingly, this study addresses the following research questions:

**RQ1.** How aware are students of each other's AI use over the course of a collaborative programming project?

**RQ2.** Is misalignment in team partners' perceptions of each other's AI use associated with team performance?

# 4 Method

## 4.1 Context

The data were collected during a collaborative programming assignment where AI use could vary between partners and is not directly observable. The assignment was part of an introductory undergraduate programming course at a large European university that enrolls students who major in subjects *other* than computer science. Data collection was designed in close collaboration with the instructors to ensure the quality of the student experience. The course comprised theory lectures, followed by a month of project work where pairs worked on the implementation of a computer game with given functionalities. During this month, three supervised in-person pair-programming sessions were hosted. These sessions took place weekly during the first three weeks of the month. Students had to attend a minimum of two of the sessions. Between the in-person sessions, students worked on their project remotely. Their contributions were shared via the version control platform GitHub. The team partners were chosen by the students themselves, and most teams were already acquainted from prior semesters. This setting allowed us to study the problem of AI-supported programming collaboration, because students worked interdependently on a shared codebase where their contributions were visible to each other but the process of producing these contributions was not. The longitudinal design repeatedly measuring peer perceptions in an authentic course, alongside the measures of prior performance and team outcome were suitable for understanding peer perceptions of AI use without disrupting the course activities.

## 4.2 Sample, Procedure and Measures

A full sample of 103 student pairs completed surveys three times within one month: before the first in-person pair-programming session (*t1*), mid-project, before session 2 (*t2*), and near the project end (*t3*). Students were offered five bonus project points for completing all three surveys. Male students comprised 65% of the sample.

**Outcome grade.** Instructors made the final project grade available for our analysis ($M = 95.9$, $SD = 5.1$; range 69–100). This grade was assigned to a team based on correctness and completeness of their implementation of the computer game.

**Prior performance**. To capture prior performance, we used the grade on a programming test taken prior to the project. It served as the prior performance covariate.

**Own AI use for the entire assignment** (measured three times). Each participant reported own AI use percentage ("What percentage of your programming involves using LLM chat or autocomplete in this project, roughly?").

**Partner AI use for the entire assignment** (measured three times***)***. Each participant estimated their team partner's AI use percentage ("What percentage of your partner's programming involves using LLM chat or autocomplete, roughly?")

**Own AI use across specific types of activities.** Students were asked to report how frequently they used AI for each type of activity in their programming project, measured only in *t2* and *t3*. The frequency for each type of activity was rated on a five-point Likert scale ranging from "never" to "always". A total of ten activities were to be selected from: automatically implementing an entire feature, generating parts of code, writing tests, writing documentation/ comments, debugging, cleaning up or restructuring code, asking if approach is correct, explaining code, and asking about principles of programming/ the programming language. These task types were adopted from categories of AI-assisted activities used in programming education research [19]. Instructors were consulted about suitability of the categories in the course. This led to the exclusion of the category "Decide type of variable" due to the programming language specificity.

**Partner AI use across specific types of activities***.* Students also estimated their partner's AI use across the same set of activities, measured only in *t2* and *t3*. These items capture understanding of the partner's AI use across programming activities, allowing to examine where estimation errors occur.

### 4.3 Analysis

**RQ1: Awareness of Partner AI Use.** To address RQ1, we analysed awareness of partner AI use at two levels: overall AI use amount and AI use type across development tasks (how much a partner relies on AI for different specific tasks). These levels capture distinct aspects of partner awareness. Slightly different methods were used to analyse these levels because amount and type pose different measurement challenges.

***AI use amount estimation*.** Students reported how much of their own amount of programming work involved AI use (self-report) and estimated their partner's amount (partner estimate) on a 0–100% scale. Since raw differences are sensitive to how individual students interpreted the percentage scale, we examined how partners perceived their *relative* AI use within the team. For each respondent we computed a normalised self–partner perception score z. z captures who the student thinks the heavier AI user is, rescaled, so it does not depend on the absolute percentages a student reports: $z = \frac{self_report - guess_about_partner}{self_report + guess_about_partner}$. Positive values indicate that a respondent perceives themselves as the heavier AI user; negative values indicate they perceive their partner as the heavier user. From this, we calculated a team-level perception misalignment score $d = z_A + z_B$ from the perceptions of team partners A and B. Summing the two partners' scores into d tests whether their perceptions are mutually consistent: if they agree on who relies more on AI, their signs offset and $d \approx 0$; if have opposing perceptions of who uses AI more than the other, $|d|$ grows. Misalignment is, therefore, not a difference in usage levels but a disagreement from the team's perceptions of relative reliance on AI. To examine if alignment changed during the collaboration, we compared absolute misalignment values ($|d|$) between consecutive waves (*t1* to *t2*; *t2* to *t3*) using Wilcoxon signed-rank tests (the data were paired and non-normally distributed).

***AI use type estimation.*** We examined students' awareness of their team partner's AI use patterns across the ten task types. For each dyad, we computed Spearman rank correlations between the partner's self-reported AI use profile and the perceiver's estimate across the ten tasks. These correlations assess if students correctly identified the tasks where their partner used AI, and how frequently.

To assess if student estimation reflects genuine partner knowledge beyond general assumptions that bias their estimate, we isolated those components, i.e. (a) assumed partner similarity and (b) normative expectation, using partial correlation. We computed partial correlations between partner self-reports and perceiver estimates while controlling for a) the perceiver's own task profile and b) all students' average AI use profile. The residual correlation measures estimation accuracy that cannot be explained by self-projection or cohort-average alone [15]. We tracked how each component changed from *t2* to *t3* as partners accumulated shared experience over time.

Further, to examine if estimates were systematically biased for particular task types, we computed the difference between partner self-reports and team partner estimates for each of the ten tasks. We tested if these differences were significant using Wilcoxon signed-rank tests and Benjamini–Hochberg (BH) correction for the false discovery rate.

**RQ2: Relationship between Perception Misalignment and Performance.** To address RQ2, we analysed the relationship between team-level perception misalignment (absolute |d|, defined above) and final project scores ($M$ = 95.9, $SD$ = 5.1; range 69–100). First, we examined bivariate associations using Spearman correlations between misalignment at each wave and project outcome to evaluate their relationship. We then fitted a robust regression model (Huber M-estimation), modelled as $\textit{project score} \sim |d|_{t1} + \textit{prior performance} + |d|_{t1} \times \textit{prior performance}$. The regression predicted project score from early misalignment (|d| at *t1*), the team's mean prior programming performance, and their interaction. Robust regression is well suited to outcome distributions exhibiting a ceiling effect, such as the one present in the project score distribution [20]. Controlling for prior performance ensured that any association between misalignment and outcomes is not confounded by prior performance that captures baseline skill differences. The interaction term tests if the performance cost of misalignment varies across prior performance levels. Simple slopes were inspected to describe this variation.

Finally, to examine whether *changes* in perception alignment during the project were associated with outcomes, we identified teams where a perception reversal occurred, i.e. a team's perception of who uses AI more has flipped. Teams with and without reversals were compared using a Mann–Whitney U test. The reversal effect was further examined using robust regression controlling for initial misalignment magnitude and prior performance.

# 5 Results

## 5.1 RQ1: How Aware Are Students of Each Other's AI Use Over the Course of the Project?

**AI use amount.** We found that students' perceptions of each other's AI use aligned over the course of the project only for 49% of the students, and alignment improved most during the first weeks. To reach this conclusion, we examined the estimation misalignment score d at each wave. At *t1*, 24% of teams showed substantial misalignment in their perceptions ($|d| > 0.2)|$. Then, alignment improved significantly over the project (Wilcoxon signed-rank tests *t1* to *t3*: $W= 2292$, $p = .003$, $r= .35$). The change from *t2* to *t3* was not significant ($p = .156$), indicating that the alignment took place in the earlier weeks. By *t3*, the proportion of substantially misaligned teams dropped to 14%. Overall, many teams improved their perceptions alignment, with 49% of teams becoming better aligned over the full project period. However, 25% of the teams became worse at judging their teammates, and 26% showed no change. This shows that although most partners' views of each other's AI use converged during collaboration, a large proportion of teams remained substantially misaligned until the end of the project.

**AI use type.** We found that students showed limited and declining accuracy in estimating their partner's AI use type pattern, but estimation error was not uniform across task types. To reach this conclusion, first, we found that students' accuracy in estimating their partners' AI use pattern was limited. To examine perceptions of partner AI use during different types of tasks, we compared the rank ordering of each partner's self-reported AI use across ten tasks with the estimates provided by their team partner. Spearman rank correlations were $r = 0.34$ at *t2* and $r = .27$ at *t3*. These low correlations indicate that variance in partners' use was not well reflected in the partner estimates.

Three categories showed significant directional bias at *t2* (Wilcoxon signed-rank tests, with BH correction): AI use for automatically implementing full features was systematically underestimated (*M*diff = +0.21, $p_adj = .049$), while AI use for asking about programming principles ($Mdiff = -0.29$, $p_adj = .029$) and for confirming the validity of an approach ($Mdiff = -0.19$, $p_adj = .049$) were overestimated. At *t3*, no task-specific differences reached significance. No other task types showed a significant directional bias at either wave, while students' absolute estimation error averaged at roughly one point on the five-point scale. We found that most of apparent awareness about a partner's AI use was a result of students' assumption that their partner uses AI the same way as them. Once this bias was removed, partner-specific accuracy was low although it increased over time.

The next step was to examine how much of the correlation between student self-report and their partners estimation of their AI use reflected genuine awareness of their partner's AI use. To understand it, we computed partial correlations controlling for two other factors: the rater's own self-report and the cohort's average AI use profile. These two controls are known biases affecting partner ratings. We found that assumed similarity was the dominant driver of partner estimates. The median correlation between a perceiver's estimates and their own self-reported usage profile was $r = .73$ at both waves. After partialling out this self-projection, the residual correlation, i.e. the component attributable to partner-specific knowledge, was substantially smaller (median partial $r = .16$ at *t2*, $r = .23$ at *t3*; all $p < .001$). The modest growth from *t2* to *t3* suggested a small gain in genuine partner knowledge over time. This gain occurred while partners' actual AI use profiles were diverging (inter-dyad profile similarity fell from median $r = .30$ at *t2* to $r = .20$ at *t3*), meaning the assumed-similarity heuristic was becoming less reliable but partners did not lose awareness what their partner was using AI for. Normative accuracy, i.e. whether a perceiver ranked the ten tasks in the same order as the population average, also explained part of overall estimation accuracy, though less than assumed similarity (median $r = 0.49$ at *t2*, $r = 0.43$ at *t3*).

In sum, what appears as awareness of a partner's AI use pattern is a product of projecting one's own behaviour and one's knowledge of what a typical student in the course does. Genuine knowledge of the specific partner is low and improves very slowly.

### 5.2 RQ2: Is Perception Misalignment Associated with Team Performance?

Teams whose members had higher misalignment in their AI use perceptions at the beginning of the project tended to receive lower scores, as captured by the Spearman correlation $\rho = -.201$ ($p = .03$) between early perception misalignment (|d| at *t1*) and the project score. Misalignment measured at later waves showed no significant association with the project score.

Robust regression (reported in Tab. 1) confirmed that early misalignment is related to outcomes, also when controlling for prior performance. A second regression including an interaction term for prior programming performance and early misalignment ($b = 0.4$, $t = 4.1$, $p < .001$) showed that prior performance was a moderator.

Table. 1. Predictors of Outcome Across Robust Linear Regression Models

| Predictor | Model 1 | Model 2 | Model 3 |
|---|---|---|---|
| Partner AI use misalignment | −4.47 (1.88)* | −7.73 (1.98)*** | −7.51 (2.07)*** |
| Prior performance | 0.03 (0.02) | 0.05 (0.02)* | 0.07 (0.023)* |
| Misalignment × Prior performance | — | 0.40 (0.10)*** | 0.423 (0.104)*** |
| Flip (*t1*→*t3* perception reversal) | — | — | −1.81* (0.838)* |
| $R^2$ | .132 | .226 | .247 |
| Adjusted $R^2$ | .117 | .206 | .217 |

**Note.** Entries are unstandardized regression coefficients (β) with standard errors in parentheses.
* p < .05. ** p < .01. *** p < .001.

Simple-slope estimates further showed that for teams one standard deviation below the mean in prior performance, increased misalignment corresponded to markedly lower outcomes ($b \approx -15.0$ per unit of |d|), whereas the association was negligible for

teams one standard deviation above the mean ($b \approx -0.32$). That is, stronger teams managed to compensate for early perceptual misalignment, weaker teams did not.

To further illustrate how both partners' perceptions jointly relate to outcomes, Fig. 1 shows a second-order polynomial response surface, estimated with heteroscedasticity-robust standard errors. X and Y axes each represent one partner's normalised self–other AI use discrepancy perception at S1: Student A = higher prior performance, Student B = lower. Color and z dimension indicate the project score fitted by the second-order polynomial model. Performance was highest near the origin, where partners are mutually aligned in their perceptions, and declined as discrepancies increased. This decline was particularly pronounced when both partners perceived the other as using less AI. Consistent with this pattern, the interaction term of student A AI use perception discrepancy and student B discrepancy was negative and significant ($b_{AB} = -121.05$, $SE$ = 49.82, $p$ = .015), whereas the remaining polynomial terms were non-significant. The model ($R^2 = .37$, $F = 10.30$) is intended here primarily as an illustrative complement to the main regression analyses, since the sample is modest for RSA, and discrepancy scores were concentrated near zero, limiting precision at the extremes.

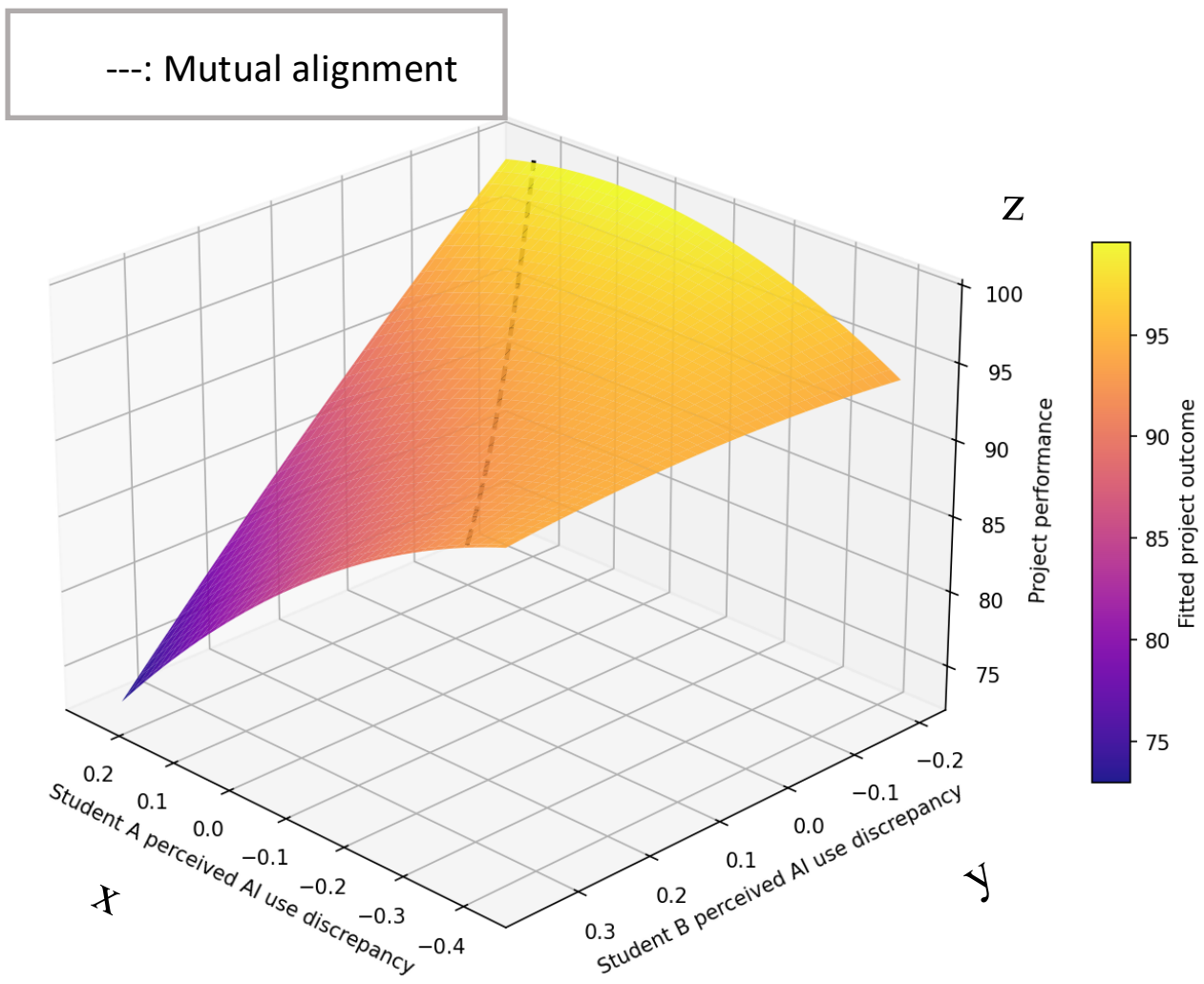

Fig. 1. Response surface of dyadic discrepancy and project outcome.

We additionally examined whether *changes* in partners' perceptions of relative AI use during the project were associated with collaboration outcomes. Teams were categorised by whether the sign of their misalignment score reversed between t1 and t3, that is, whether the team's view of who uses more AI flipped. Teams experiencing a direction reversal achieved lower scores than teams with stable perceptions (medians: 94.5 vs. 98 percentage points; Mann–Whitney $U = 192$, $p = .011$, $r_rb = .427$). A robust regression (see Tab. 1) controlling for initial misalignment and prior performance yielded a consistent pattern: direction reversals remained associated with lower outcomes ($b = -1.90$, $t = -2.32$, $p = .021$). However, only 25 teams experienced a reversal, so these results should be treated as indicative of the relevance of stark changes in AI use perception rather than conclusive.

## 6 Discussion

Regarding RQ1, we find that awareness of partners' AI use is limited and largely explained by heuristic inference processes such as assumed similarity rather than partner-specific knowledge. Apparent accuracy in estimating partners' AI-use patterns was moderate, but this was substantially reduced after accounting for assumed similarity. This suggests that ongoing collaboration does not provide sufficient partner-specific information to substantially correct these self-based inferences. Assumed similarity effects may be pronounced in our sample since many participants have prior relationships. The prior relationship is is known to affect assumed similarity, according to literature showing that partners project their own habits onto someone they feel they know well [21]. Consistent with this, qualitative reports where participants in AI-supported programming assumed that others use AI in ways similar to themselves [22]. Students' biases in estimations about their partners' AI use furthermore suggest that judgments are structured by norms of acceptable AI usage. The biases students showed in their estimation were underestimation of their partners' AI use for fully automated feature implementation and overestimation of partners' AI use for questions on programming principles or validation of a solution strategy. This aligns with literature showing that judgments are shaped by expectations about what behaviours are typical or appropriate in a given context [16].

Over time, the increase in correlation of self-report and partner estimation of AI use after partialling out similarity bias or normative bias suggests a modest increase in partner-specific knowledge. Students' partner understanding beyond reliance on assumed similarity heuristics is particularly important given team members' AI-use patterns appear to diverge at the later stages. Divergence of AI use profiles may occur naturally over a project, as students take on more specialised roles, such as writing tests. Although pair-programming sessions as applied in the course are often assumed to support shared mental models, we do not observe a consistent improvement in partner-specific accuracy over time. One possible explanation is that AI use differs between individual work and in-person collaboration, such that opportunities to observe a partner's AI use remain limited even during joint sessions. The limited increase in partner knowledge may also be related to the course setting, where most students were previously acquainted from other courses together and had selected their partner themselves based on prior relationships. The familiarity between partners is also likely to inflate similarity assumption bias, as prior research suggests [21].

Regarding RQ2, we find that misalignment in partners' perceptions of AI use is associated with team performance. In particular, greater misalignment at the beginning of the project was associated with lower final project scores. This pattern is consistent with prior research on collaborative work where accurate beliefs about a partner's behaviour have been shown to support effective coordination. Later partner alignment in AI use perceptions does not appear to offset the association between early misalignment and performance. The concentration of the performance effect at *t1* suggests that early perception gaps may be particularly consequential during initial coordination phases, when teams establish task allocation, expectations, and working strategies. To find plausible pathways of how misalignment in partners' perceptions of AI use relates to collaboration performance likely relevant mechanisms can be identified from prior research. Research has found several concrete mechanisms to be relevant to team effectiveness in students' programming collaboration: accurately monitoring mutual performance, backup behaviour and mutual trust (a belief that team members do their best) [23]. Further investigations are warranted to understand the influence of AI on these factors of collaboration.

The relationship between misalignment and outcomes further depended on individuals' prior performance. The negative association between misalignment and performance was substantially stronger for teams with lower individual prior performance, while it was negligible for teams with high prior performance. This pattern suggests that stronger teams may be less dependent on accurate perceptions of AI use for effective coordination, potentially because they can rely more on shared technical understanding or compensate for misaligned expectations. When students have less prior programming skills, inaccurate beliefs about a partner's AI reliance may be especially costly because they may also be less able to judge the quality, correctness, or limitations of AI-supported code independently. This interpretation is consistent with recent evidence that novices accept buggy or difficult-to-understand AI-generated solutions, often report not understanding why AI suggestions work, and have particular difficulty correcting AI-generated code once problems arise [24].

Additional analyses of perceptions showed that when perceptions of which partner used more AI changed between the beginning and the end of the project, the team tended to achieve lower outcomes than teams with stable perceptions, even when controlling for initial misalignment and prior performance. Rather than simply indicating that partners learned over time, such changes may signal that early assumptions were revised only after collaboration difficulties had already emerged.

The findings point to a structural feature of AI-supported collaboration: when a coordination-relevant behaviour is not directly observable, partners rely on heuristic inferences that produce only partially accurate beliefs. This suggests that AI-use transparency may be relevant not only for academic integrity but also as a coordination resource. Designing learning analytics and pedagogical scaffolds to make AI use more visible or interpretable to collaborators may therefore support more accurate partner models. Recent work on shared and visible human–human–AI settings suggests that when AI assistance is visible to a peer, participants feel more responsibility to understand suggestions before adopting them [13], [25]. In the present context, this raises the possibility that lightweight mechanisms such as shared prompt histories, AI-use annotations, or peer-visible records of AI-supported work could improve partner understanding by making otherwise private AI use more accountable and interpretable.

Whether such interventions would improve coordination in practice remains an open empirical question.

Effective intervention requires understanding more precisely where awareness of a partner's AI use is most consequential, and whether pedagogically incentivised disclosures would improve outcomes. Furthermore, it is yet to be identified what the appropriate awareness signal would be, such as usage-type-level AI-use disclosure or analytics, or code level signals integrated into the shared repository. These are empirical questions that require both improved measurement and an experimental evaluation of designed interventions. Our findings strongly suggest that the problem of perceiving partner AI use accurately is consequential and further work is warranted.

Our study has several limitations. First, the design is observational, so the reported associations should not be interpreted causally. Second, the measures of AI use were based on self-reports and partner estimates rather than behavioral traces, and are therefore subject to memory error, scale interpretation differences, and possible social desirability effects. Additionally, the outcome grade showed a ceiling effect, which affects modeling options and may have led to underestimation of the associations. Third, the study did not analyse the mechanisms through which perception misalignment may affect collaboration, such as trust calibration, code review behaviour, task allocation, or communication patterns. This remains to be studied. Fourth, the observed effect sizes were small to moderate. This finding is unsurprising given that collaborative project outcomes are shaped by many factors beyond AI-use perception. Finally, the study was conducted in a single course context with students who were often already familiar with their partner. This context may limit generalizability to other educational settings, team compositions, or project formats.

## 7 Conclusion

When each member of a programming team independently uses AI, the human-to-human coordination between them is affected in ways that are consequential. Teams whose members hold inaccurate beliefs about each other's AI reliance from the project outset perform worse, especially when baseline individual prior performance was low and coordination errors are most costly. Most of what looks like knowledge of the partner is a projection of own patterns of AI use, and partner-specific understanding only slightly improves over time. Structured in-person collaboration appears insufficient to close these gaps despite its potential for shared understanding. These findings reframe AI transparency beyond that of an academic integrity concern towards a coordination resource and a call for examining the mechanisms that can help partners coordinate effectively. The findings suggest that AI-use transparency features or prompting disclosure could be integrated into collaborative programming in early phases of remote collaborative work. Awareness support should particularly target teams with lower prior performance, where misaligned perceptions carry the highest performance cost.